\documentclass[aps,prb,article,twocolumn,showpacs,preprintnumbers,amsmath,amssymb,superscriptaddress]{revtex4}
\date{\today}
\usepackage{epsfig}
\usepackage{subfigure}
\usepackage{graphicx}
\usepackage{dcolumn}
\usepackage{bm}
\usepackage{float}
\usepackage{comment}
\usepackage{color}
\usepackage{ulem}
\usepackage[textsize=footnotesize,textwidth=0.5in]{todonotes}

\newcommand{\tJ}{{$t$-$J$}}

\newcommand{\tJl}{{$t$-$J$-like}}
\newcommand{\ttprJ}{{$t$-$t^\prime$-$J$}}
\newcommand{\tJs}{{$t$-$J$-$3s$}}
\newcommand{\ttprJs}{{$t$-$t^\prime$-$J$-$3s$}}

\begin{document}

\title{Numerical Investigation of Spin Excitations in a Doped Spin Chain}

\author{Ekaterina M. P{\"a}rschke}\thanks{E. M. P{\"a}rschke and Y. Wang contributed equally to this work.}
\affiliation{Department of Physics, University of Alabama at Birmingham, Birmingham, Alabama 35294, USA}

\author{Yao Wang}\thanks{E. M. P{\"a}rschke and Y. Wang contributed equally to this work.}
\affiliation{Department of Physics, Harvard University, Cambridge 02138, USA}

\author{Brian Moritz} \affiliation{Stanford Institute for Materials and Energy Sciences, SLAC National Accelerator Laboratory and Stanford University,
Menlo Park, California 94025, USA}

\author{Thomas P. Devereaux}
 \affiliation{Stanford Institute for Materials and Energy Sciences, SLAC National Accelerator Laboratory and Stanford University,
Menlo Park, California 94025, USA}
 \affiliation{Department of Materials Science and Engineering, Stanford Universoty, Stanford, California 94305, USA}
 
\author{Cheng-Chien Chen}
 \affiliation{Department of Physics, University of Alabama at Birmingham, Birmingham, Alabama 35294, USA}

\author{Krzysztof Wohlfeld}
\affiliation{Institute of Theoretical Physics, Faculty of Physics, University of Warsaw, Pasteura 5, PL-02093 Warsaw, Poland}

\date{\today}
\begin{abstract}
We study the doping evolution of spin excitations in a 1D Hubbard model and its downfolded spin Hamiltonians, by using exact diagonalization combined with cluster perturbation theory. In all models, we observe hardening (softening) of spin excitations upon electron (hole) doping, which are reminiscent of recent experiments on 2D cuprate materials. We also find that the 3-site and even higher-order terms are crucial for the low-energy effective spin models to reproduce the magnetic spectra of doped Hubbard systems at a quantitative level. To interpret the numerical results, we further employ a strong coupling slave-boson mean-field theory. The mean-field theory provides an intuitive understanding of the overall compact support of dynamic spin structure factors, including the shift of zero-energy modes and change of spin excitation bandwidth with doping. Our results can serve as predictive benchmarks for future inelastic x-ray or neutron scattering experiments on doped 1D antiferromagnetic Mott insulators.
\end{abstract}



\maketitle

\section{Introduction} 
Among the intertwined degrees of freedom in correlated materials, spin and spin excitations are of particular significance. Various emergent phases, including pseudogap, stripe, high-$T_c$ superconductivity and strange metal, arise from doping a spin-ordered Mott insulator.~\cite{kivelson2003detect,keimer2015quantum} Spin fluctuations are widely believed to act as the pairing glue for unconventional superconductivity.~\cite{tsuei2000pairing, scalapino2012common, maier2016pairing} In addition, a number of novel solid-state phenomena are driven by the spin interaction, including magnetoresistance,~\cite{camley1989theory} spin liquid,~\cite{balents2010spin} and topological insulator.~\cite{qi2011topological}
Spin excitations also offer a promising route for the development of future magnetic, spintronics and transistor devices.~\cite{wolf2001spintronics, vzutic2004spintronics} The importance of characterizing spin excitations has dramatically pushed down the experimental resolution of the resonant inelastic x-ray scattering (RIXS) and inelastic neutron scattering (INS) experiments.~\cite{braicovich2009dispersion,schlappa2009collective,le2011intense,schlappa2012spin,dean2013persistence,GuochuDengPRB2018,INOSOV2016} 
In contrast, theoretical description of spin excitations in correlated systems remains an open question, except for a few simplified toy models.~\cite{wang2018theoretical}

Spin excitations are not even fully understood in a one-dimensional (1D) correlated system, which directly describes Sr$_2$CuO$_3$, SrCuO$_2$,~\cite{fujisawa1999angle, neudert1998manifestation,kim1996observation} and organic salts such as TTF-TCNQ.~\cite{claessen2002spectroscopic}
Very recently, 1D systems have been also realized using cold atoms~\cite{YangGri2018} and quantum circuits.~\cite{Anthore2018} The minimal physics of correlated electrons can be described by the Hubbard model. An asymptotically exact solution of the ground-state energy of 1D half-filled Hubbard model was obtained by Lieb and Wu by mapping the Hubbard model onto algebraic equations through the Bethe ansatz.~\cite{lieb1968absence,voit1995one} Though exact, its implicit wavefunctions lack direct intuitions and remain unhandy in calculating the excitation spectra relevant to experiments.~\cite{Essler} Different aspects of 1D half-filled Hubbard model have been extensively studied using various numeric techniques,~\cite{Berciu2000, Knizia2012, Rodriguez2014,LiuWang2015,Booth2015} and the density matrix renormalization group (DMRG)\cite{white1992density, white1993density, schollwock2005density,schollwock2011density} is believed to provide an exact solution of the ground-state properties.  

Compared to the ground-state properties, the excitation or dynamics is much more difficult to compute theoretically. At half filling, the spin spectrum can be solved using Takahashi's theory,~\cite{ebetaler1994scattering,essler19942} while with doped carriers the spectrum can become complicated to obtain due to the breaking of SO(4) symmetry.~\cite{deguchi2000thermodynamics} In realistic materials, the existence of next-nearest-neighbor hopping and the particle-hole asymmetric nature brings extra complexity even at half filling. Therefore, obtaining a comprehensive understanding of spin excitations in a doped 1D Hubbard model remains a challenge.~\cite{preuss1994spectral, Nishimoto2008, Rodriguez2014,kung2017}

Due to the effective spin-charge separation nature in 1D,~\cite{preuss1994spectral,kim2006bj} the elementary excitations of the 1D Hubbard model are spinon and holon carrying respectively only spin and charge degrees of freedom, and they propagate with different velocities.~\cite{giamarchi2004quantum,voit1995one}
Attempts have been made to simplify the Hubbard model and obtain a low-energy Hamiltonian.
Using the canonical perturbation theory one can show that the Hubbard model can be mapped onto a \tJ\--like model~\cite{chao1977ka, chao1978canonical}. The latter class of models requires far less numerical effort due to a restricted Hilbert space,~\cite{giamarchi2004quantum} 
and provides a good approximation for correlated electrons. In particular, in the half-filled limit the \tJ\ model becomes identical with the Heisenberg model. In this case, the ground state of such an $S=1/2$ spin 1D
Heisenberg model can be obtained using the Bethe ansatz~\cite{Bethe1931} and exhibits a quasi-long-range antiferromagnetic (AFM) order.
A few conclusions have been drawn analytically, including the des Cloizeaux-Pearson lower bound of the excitation continuum~\cite{desCloizeaux1962} and the elementary spinon excitation.~\cite{Faddeev1981}  
The dynamical spin structure factor of a half-filled spin chain has been calculated with both (analytical) Bethe ansatz~\cite{Caux2006, Kohno2009, Klauser2011} and (numerical) DMRG techniques,~\cite{Hallberg1995, Kuehner1999} and the results are consistent with the INS and RIXS experiments.~\cite{Walters2009, schlappa2012spin, Lake2013, Mourigal2013, Schlappa2018} 
However, away from half filling, due to the unphysical restriction of Bethe ansatz~\cite{Ha1994} and general difficulty of treating dynamical properties in numerics, a systematic description of the doping evolution of spin excitations is still lacking.~\cite{Deisz1992, Tohyama1995, Zhang1997, Forte2011, Kumar2018}
 
Another motivation for this work is the recent discovery of persistent spin excitations in doped (2D) cuprate high-$T_c$ materials.~\cite{le2011intense,dean2013persistence,LeeNaturePhys2014,ishii2014high,Ellis2015} 
In these experiments, paramagnon excitations were observed along the antinodal direction even in the overdoped regime. This lack of softening upon hole doping is quite puzzling. In addition, the paramagnons in electron-doped cuprates were found to harden with doping, which is unexpected considering the loss of long-range AFM order. Theoretical explanations for these phenomena include the attribution to the 3-site correlated hopping~\cite{jia2014persistent} and the longer-range hopping $t^\prime$.~\cite{wang2014real,LiLinPRB2016} Without 
efficient and reliable numerical tools in 2D correlated systems, it is hard to reach a definitive answer.
Therefore, an analog analysis in a 1D system is on demand as a key step.~\cite{kung2017, Tohyama2018}

In this paper, we systematically study the doping evolution of the spin dynamical structure factor of the 1D Hubbard, \tJ\, and \tJs\ models (see Sec.~\ref{sec:model} for the model definitions) using exact diagonalization (ED) and cluster perturbation theory (CPT). In the absence of
longer-range hopping $t^\prime$, we find that the \tJl\ models can well capture spin excitations of the Hubbard model, providing lower and upper bounds of the compact support. 
However, for a finite $t^\prime$, the quantitative agreement between the Hubbard and \tJl\ models becomes worse.
We also find that the hardening of spin excitations upon electron doping and their persistence upon hole doping are well-addressed by the presence of $t^\prime$, with an additional contribution from the 3-site term.
We further supplement these results with a simplified analytical slave-boson mean-field approach, which allows for an intuitive interpretation of the numerical results. Our systematic analysis not only provides quantitative predictions for INS and RIXS experiments on 1D materials, but also addresses the anomalous hardening phenomenon in 2D cuprates.

This article is organized as follows. In Sec.~\ref{sec:model}, we introduce the Hubbard model and its low-energy effective Hamiltonians, including the \tJ\ and \tJs\ models. In Sec.~\ref{sec:methods}, we introduce two approaches to evaluate the dynamical spin structure factor, a numerical CPT and an analytical slave-boson mean-field approach. We present the calculated spectral results in Sec.~\ref{sec:results} and the discussions in Sec.~\ref{sec:discussions}. We conclude the study in Sec.~\ref{sec:conclusions}.

\section{Models}\label{sec:model}
The single-band Hubbard model is the simplest model describing strong Coulomb interactions between electrons,~\cite{anderson1987resonating,zhang1988effective} and the Hamiltonian reads
\begin{align} \label{Hubbard_single-band}
      \mathcal{H}_\mathrm{Hubbard}=
-\!\sum_{i,j,\sigma}\! t_{ij} \! \! \left( {c}^\dagger_{i \sigma}{c}^{\phantom{\dagger}}_{j \sigma}+\mathrm{h.c.}\right)\!  +U\sum_{i}{n_{i\uparrow}n_{i\downarrow}}.
\end{align}
Here, $c^\dagger_{i \sigma}$ denotes an operator creating an electron on site $i$ carrying spin $s_z\equiv \sigma=\pm1/2$, $n_{i \sigma}=c^\dagger_{i \sigma}c^{\phantom{\dagger}}_{i \sigma}$ is the electron number operator, and $t_{ij}$ is the hopping integral.

The hopping integrals usually decay quite rapidly with the interatomic distance $|i-j|$ for the Wannier wavefunctions. In the models for quasi-2D cuprates, they are typically truncated at the nearest-neighbor hopping $t$, the next-nearest-neighbor hopping $t^\prime$, or (at most) the third neighbor hopping $t''$.~\cite{Tohyama2000, Damascelli2003} The latter two hoppings break particle-hole symmetry, as is seen in the well-known asymmetric phase diagram of the cuprates. The relatively narrow AFM phase in hole-doped cuprates generally requires the leading longer-range hopping to be negative. Therefore, to be consistent with 2D cuprate system, we consider a negative next-nearest-neighbor hopping $-0.3\leq t^\prime\leq 0$ throughout this paper.

In the strong-coupling limit $U\gg t$, the single-band Hubbard model (with $t^\prime=0$) can be mapped onto the \tJ\ model~\cite{chao1977ka, chao1978canonical,belinicher1994consistent,belinicher1994range,stephan1992single} by projecting out doubly occupied states:~\cite{schrieffer1966relation}
\begin{equation}\label{tJmodel_single-band}
      \mathcal{H}_\textit{t-J}=- \!\sum_{ i,j,\sigma}\!  t_{ij} \!\left(\tilde{c}^\dagger_{i \sigma}\tilde{c}^{\phantom{\dagger}}_{j \sigma}+\mathrm{h.c.}\right)\!+
      J\sum\limits_{\langle i,j\rangle}{\left(\mathbf{S}_{i}\!\cdot\!\mathbf{S}_{j}-\frac{\tilde{n}_{i}\tilde{n}_{j}}{4}\right)},
\end{equation}
where the spin-exchange energy $J=4t^2/U$.  
In contrast to the fermionic operators representing real electrons in the Hubbard model, creation (annihilation) operators $\tilde{c}^\dagger_{i \sigma}$ ($\tilde{c}_{i \sigma}$) here act in the restricted Hilbert space where double occupancy is forbidden: $\tilde{c}^\dagger_{i \sigma} = c^\dagger_{i \sigma} (1-n_{i\bar{\sigma}})$ and $\tilde{n}_{i \sigma}=n_{i\sigma}(1-n_{i\bar{\sigma}})$. The spin operator $\textbf{S}_i\cdot\textbf{S}_j\! =\! S_i^zS^z_j\!+\!\left(S_i^+S_j^-+S_i^-S_j^+\right)/2$, with $S_i^z\!=\!(n_{i \uparrow}\!-\!n_{i\downarrow})/2$ and $S_i^+\!=\!(S_i^-)^\dagger\!=\!\tilde{c}^\dagger_{i\uparrow} \tilde{c}^{\phantom{\dagger}}_{i\downarrow}$. By performing a particle-hole transformation, one can map an electron-doped system to a hole-doped one with only a sign change of the longer-range hoppings $t^\prime$ and $t^{\prime\prime}$.~\cite{LeePRL2013}
We adopt this trick for the analysis of electron-doped system in the \tJ\  framework, but still define the physical parameters including $t^\prime$, doping concentration $x$, electron density $n$, Fermi energy $E_F$, and Fermi momentum $k_F$ in the original basis, consistent with the Hubbard model. 
 
During the strong-coupling expansion, in principle Eq.~\eqref{tJmodel_single-band} should contain one more term of the same order as the spin exchange interaction ($t^2/U$). This so-called 3-site term~\cite{stephan1992single,jefferson1992derivation,spalek1988effect,von1990single,eskes1994h,belinicher1996generalized,eskes1996hubbard,kuz2014comparison} is described by
\begin{equation} \label{tJH}
\mathcal{H}_{3s}= - \frac{J}{4}\!\!\sum_{\langle i,j\rangle,\langle i,j^\prime\rangle\atop j\neq j^\prime,\sigma}\! \! \left(\tilde{c}^\dagger_{j^\prime\sigma}\tilde{n}^{\phantom{\dagger}}_{i\bar{\sigma}}\tilde{c}^{\phantom{\dagger}}_{j\sigma}\!- \! \tilde{c}^\dagger_{j^\prime\sigma} \tilde{c}^\dagger_{i\bar{\sigma}} \tilde{c}^{\phantom{\dagger}}_{i\sigma} \tilde{c}_{j\bar{\sigma}}\right).
\end{equation}
Adding it to the \tJ\ model produces the correct second-order expansion of the Hubbard model. The \tJ+3-site Hamiltonian (denoted as \tJs\ in the following) reads:
\begin{align}
\mathcal{H}_{t-J-3s} = \mathcal{H}_{t-J} +\mathcal{H}_{3s}.
\end{align}
Since the 3-site terms do not contribute to the ground state at half filling, they have quite often been ignored. However, they were shown to be important for describing the single-particle excited states and accordingly the motion of the hole introduced in the ground states with magnetic correlations.~\cite{wang2015origin, wang2018influence}

As the 3-site term describes an effective long-range hopping mechanism within the same AFM sublattice, it favors the persistence of spin correlations in the doped \tJl\ models. Indeed, it was proposed to partially account for the anomalous paramagnon hardening in the electron-doped 2D Hubbard model.~\cite{jia2014persistent}
Note that the above expressions for the spin exchange $J$ and 3-site terms are precise only for $t' \ll t$,~\cite{Bala1995} beyond which limit the longer-range spin exchange $J^\prime$ and other 3-site terms $\propto t^\prime t/U$ or $\propto t^{\prime2}/U$ are non-negligible. In 2D cuprate materials, it has been pointed out that these effective longer-range terms are important to explain \textit{e.g.}~the asymmetric nodal and antinodal magnon excitations.~\cite{peng2017influence,wang2018magnon} We will discuss the contribution from $t'$ in Sec.~\ref{sec:discussions}.

\section{Methods}\label{sec:methods}
To understand the doping evolution of spin excitations in correlated systems, we examine the zero-temperature spectral properties of the Hubbard, \tJ\, and \tJs\ models, characterized by the dynamical spin structure factor
\begin{equation} \label{Sqw}
 S(q,\omega) = \frac{1}{\pi}{\mathrm{Im}\langle G|s_{-q}\frac{1}{\mathcal{H} - \omega-E_G-i\delta}s_q|G\rangle},
\end{equation}
where $|G\rangle$ is the ground state of the model Hamiltonian with energy $E_G$. Fourier transformation of the spin flip operator is written as $s_q=\sum_j e^{iqj}S_j^+/\sqrt{L}$, where $L$ is the number of lattice sites. 

In order to evaluate $S(q,\omega)$, we employ two complementary approaches. We first compute $S(q,\omega)$ numerically using finite-size cluster ED, with an infinite-chain extrapolation in the CPT manner [see Sec.~\ref{sec:EDCPT}]. This approach treats all many-body states of a small cluster exactly and is applicable for any tight-binding models. Such extrapolation is approximate but well-controlled, where the result can be systematically improved by increasing the cluster size in the ED calculations. 
This approach provides a quantitative description for the evolution of spin excitations upon doping the Hubbard or \tJl\ models. 
The second approach we adopt is an analytical slave-boson mean-field theory [see Sec.~\ref{method:MFT}], which decouples the interacting degrees of freedom in \tJl\ models to evaluate the $S(q,\omega)$ compact support (regions of allowed excitations in the energy-momentum space). The latter analytical approach is also approximate but not limited by the system size, and it can help to interpret the numerical spectra more intuitively.

\subsection{Exact Diagonalization and Cluster Perturbation Theory}\label{sec:EDCPT}

Exact diagonalization (ED) can treat interaction effects exactly and provide direct access to the many-body wavefunctions.
Since it usually requires the enumeration of the full Hilbert space of a many-body system, which scales exponentially with the system size, the  
calculation is usually restricted to a finite-size cluster.  
The excited states, however, can be nevertheless captured properly by linking finite clusters into an effectively infinite system.~\cite{maier2005quantum} One successful approach is cluster perturbation theory (CPT), which treats the Hamiltonian inter-cluster terms  
as perturbation and extrapolates the result on a finite cluster to that of an infinite system.~\cite{Senechal2000,Senechal2002,Eder2005} It has been demonstrated that CPT can solve the single-particle spectral function for Hubbard and \tJl\ models very well for the 1D and 2D hypercubic lattices.~\cite{wang2015origin, kohno2015spectral, wang2018influence} This approach was recently extended to the spin spectrum.~\cite{Chen2015PRB, kung2017}  

Within the ED+CPT framework, one 
first evaluates the ground state of a given Hamiltonian in an open-boundary cluster. Here we adopt the parallel Arnoldi approach~\cite{lehoucq1998arpack} with the Paradeisos implementation~\cite{jia2018paradeisos} performed on a $16$-site chain employing continued fraction expansion (CFE).~\cite{dagotto1994correlated} 
A broadening $\delta=0.15t$ is chosen to account for the intrinsic and experimental broadening (of e.g. a typical RIXS experiment~\cite{braicovich2009dispersion,schlappa2009collective,le2011intense,schlappa2012spin,dean2013persistence}) in the numerical evaluation of Eq.~\eqref{Sqw}. 
In 1D systems, the vertex correction describing inter-cluster interaction has been shown to be less important due to a minimal boundary-bulk ratio,~\cite{Chen2015PRB, kung2017} 
and it is set to zero for convenience in this work. Moreover, the inter-cluster hopping does not enter the CPT vertex, because the Green's function is replaced 
by the spin correlation function on the two-particle level. A precise treatment of the hopping-induced vertex correction would require a generalization into a four-point Green's function, which is substantially more complicated.

\subsection{Slave-boson mean-field approach}
\label{method:MFT}
To gain a more intuitive understanding of the spin dynamics in the 1D \tJ\ model, we also employ an analytical slave-boson mean-field approach. This technique utilizing the random phase approximation (RPA) has been successfully used to describe the structure above the pair-breaking peak in the Raman scattering in HgBa$_{2}$CuO$_{4+\delta}$,~\cite{Zeyher2013} and explain the large hardening effect in the electron-doped cuprates as discussed in Refs.~\onlinecite{LiLinPRB2016,LeeNaturePhys2014}. 

Here we derive the slave-boson equations for a hole-doped system, and the electron-doped case can be described by reversing the signs of $t^\prime$ and $x$ and by transforming Fermi momentum $k_F$ to $\pi-k_F$.~\cite{LeePRL2013}
Within the slave-boson framework, the electron creation operator is expressed in terms of two operators 
\begin{align}
\tilde{c}^\dag_{i \sigma}=f^\dag_{i \sigma}a_{i}, 
\end{align}
where $f^\dag_{i \sigma}$ creates a fermion carrying the spin degree of freedom (spinon), while $a_i$ annihilates a boson carrying the charge degree of freedom (holon).~\cite{Raczkowski2013} 
The kinetic energy term of the \tJ\ model is then written as
\begin{eqnarray}\label{eq:sb-kinetic}
       \mathcal{H}_\textit{t} = \sum_{\langle i,j\rangle,\sigma} t_{ij} \tilde{c}^\dagger_{i\sigma}\tilde{c}^{\phantom{\dagger}}_{j\sigma} &=& \sum_{\langle i,j\rangle,\sigma} f_{i\sigma}^\dagger a^{\phantom{\dagger}}_i  a_j^\dagger f^{\phantom{\dagger}}_{j\sigma}.
\end{eqnarray}

To exclude double occupancies, a constraint 
\begin{align} \label{eq:constraint}
a_i^\dagger a^{\phantom{\dagger}}_i + \sum_\sigma f^\dag_{i \sigma}f^{\phantom{\dagger}}_{i \sigma} \equiv 1
\end{align}
has to be imposed for every site $i$.~\cite{Yuan2004} We obtain for the charge operator $\tilde{n}_i = \sum_\sigma f_{i\sigma}^\dagger a^{\phantom{\dagger}}_i a_i^\dagger f^{\phantom{\dagger}}_{i\sigma} = 1- a_i^\dagger a^{\phantom{\dagger}}_i $ and for the spin operator $\mathbf{S}_{i} = \frac12 \sum_{\sigma,\sigma^\prime} f_{i\sigma}^\dagger \boldsymbol{\sigma}_{\sigma,\sigma^\prime}f^{\phantom{\dagger}}_{i\sigma^\prime}$. Therefore
\begin{align}\label{app:MFT2}
    \mathcal{H}_\textit{J} = & 
    -\frac12 J \sum_{\langle i,j\rangle}\sum_{\sigma_1,\sigma_2} f_{i\sigma_1}^\dagger f^{\phantom{\dagger}}_{j\sigma_1}f_{j\sigma_2}^\dagger f^{\phantom{\dagger}}_{i\sigma_2}  \nonumber \\   
    &- \frac12 J \sum_{\langle i,j\rangle}\left(1-a_i^\dagger a^{\phantom{\dagger}}_i\right)\left(1-a_j^\dagger a^{\phantom{\dagger}}_j\right).
\end{align}

We then perform a mean-field decoupling of spinons and holons. To this end, we define the resonance valence bond mean fields $D = \sum_\sigma \langle f_{i\sigma}^\dagger f^{\phantom{\dagger}}_{i+1,\sigma}\rangle$ and $D^\prime = \sum_\sigma \langle f_{i\sigma}^\dagger f^{\phantom{\dagger}}_{i+2,\sigma}\rangle$, reflecting the strengths of nearest and next-nearest neighbor spin bonds, as well as the mean-field holon kinetic energies $G = \langle a_i^\dagger a^{\phantom{\dagger}}_{i+1}\rangle$ and $G^\prime = \langle a_i^\dagger a^{\phantom{\dagger}}_{i+2}\rangle$. As we are interested in the gapless spin excitations, we do not consider possible spinon-pairing term that can potentially lower the mean-field energy but usually gaps out the spinons.
We note that the values of $D$, $D'$, $G$, and $G'$ are real due to the inversion symmetry. With the mean-field decoupling, the \tJ\  Hamiltonian is expressed as
\begin{eqnarray}\label{eq:tJMFT}
    \mathcal{H}_{t-J}^{\rm MFT} = \sum_{k\sigma} \xi_k f_{k\sigma}^\dagger f^{\phantom{\dagger}}_{k\sigma} + \sum_k \zeta_q a_q^\dagger a^{\phantom{\dagger}}_q,
\end{eqnarray}
where $f_{k\sigma} = \sum_k e^{-ikj} f_{j\sigma}\sqrt{L}$ and $a_q = \sum_q e^{-iqj} a_{j}\sqrt{L}$ are the mean-field free spinon and holon annihilation operators in momentum space.
The spinon and holon dispersion relations read 
\begin{align}\label{xik}
    \xi_k &= -(JD + 2tG)\cos k - 2t^\prime G^\prime \cos 2k,\\
    \label{qk} \zeta_q &= (1-|x|)J - (JG + 2tD)\cos q - 2t^\prime D^\prime \cos 2q.
\end{align}
Here, $x$ defines the doping level away from half filling (with $n=1+x$ being the average number of electrons per site),
and constant terms are neglected in Eq.~\eqref{eq:tJMFT}.

To determine the values of the mean fields, we first assume that the translational symmetry is preserved and that $D>0$ (will be demonstrated later). Then, it can be calculated as an average over bonds: $ D = \sum_{i\sigma} \langle f_{i\sigma}^\dagger f_{i+1,\sigma}\rangle/N = {2\sin (k_F)}/{\pi}$. Similarly, $D^\prime = \sin (2k_F)/\pi$. Here, the Fermi momentum $k_F = \pi n/2$, {which for a hole-doped system amounts to} $k_F = \pi(1-|x|)/2$. In a similar manner we obtain the mean-field holon kinetic energies $G  = \sum_i \langle a_i^\dagger a_{i+1}\rangle/N = {\sin (q_F)}/{\pi} $ and $G^\prime = {\sin (2q_F)}/ ({2\pi})$, with $q_F = |x|\pi$.

The mean-field solution employing these values provides the correct trend for the dynamical structure factor upon doping, but it underestimates the correlation effects. Since the solution of the undoped spin chain is known exactly,~\cite{Klauser2011} we can match our mean-field solution with the exact undoped spinon bandwidth, by taking $D = (\pi \sin k_F) /2$. Also, we notice that 
$G^\prime \approx G$ for small $x$, and we set them equal in this study $G^\prime \equiv G$. As shown later, these assumptions result in good agreement with the numerical solution.
Upon substituting the Fermi momenta into Eq.~\eqref{xik}, the analytical form of the spinon dispersion reads:
\begin{eqnarray}\label{eq:xik}
    \xi_k\! =\! -\!\left[\frac{\pi J}{2}\!\cos\! \frac{\pi x}{2}\!+\!2t\frac{\sin(|x|\pi)}{\pi}\right]\!\cos k  - 2t^\prime\frac{\sin(|x|\pi)}{\pi} \cos2k.\nonumber\\
\end{eqnarray}
A similar dispersion has been obtained in Ref.~\onlinecite{LiLinPRB2016} for a 2D spin model, where the $G$ and $G^\prime$ were taken to be half of the Gutzwiller renormalization factor $\sim |x|/(1+|x|)$. Our $G$ factor is very close to that value for $|x|<0.5$. 

After the mean-field decoupling, the Hamiltonian Eq.~\eqref{eq:tJMFT} can be treated as a non-interacting tight-binding model.
Spin susceptibility can therefore be calculated using the Lindhard theory:~\cite{mahan2013many,giamarchi2004quantum}
\begin{equation} \label{eq:lindhard}
   \chi(q,\omega)=\frac{1}{4N}\sum\limits_{k\in BZ}{\frac{f_F(\xi_k)-f_F(\xi_{k+q})}{\omega+\xi(k)-\xi(k+q)+i\delta}},
\end{equation}
where the Fermi distribution at zero temperature $f_F(k)=\theta(k_F-k)$ is a Heaviside function, and $\delta=0^+$.
The spin dynamical structure factor is then computed using $S(q,\omega)=-\textrm{Im}\chi(q,\omega) / \pi$.

It will be shown below that the spin spectra of the 1D Hubbard or \tJl\ models can be qualitatively reproduced by this simple slave-boson mean-field approach.
As this approach requires a decoupling of spin and charge degrees of freedom, such a good agreement between the slave-boson mean-field and the numerically exact results 
can be viewed as a signature of the spin-charge separation~\cite{lieb1968absence}, which occurs even at finite doping levels.~\cite{Imambekov2012} 

\section{Results}\label{sec:results}
Using the two approaches described in Sec.~\ref{sec:methods}, we calculate spin spectra of the Hubbard, \tJ\, and \tJs\ models. In order to highlight the role played by the particle-hole symmetry, we present the dynamical spin structure factor $S(q,\omega)$ for $t^\prime = 0$ and finite $t^\prime$ separately.
\begin{figure*}[!t]
\begin{center}
\includegraphics[width=2\columnwidth]{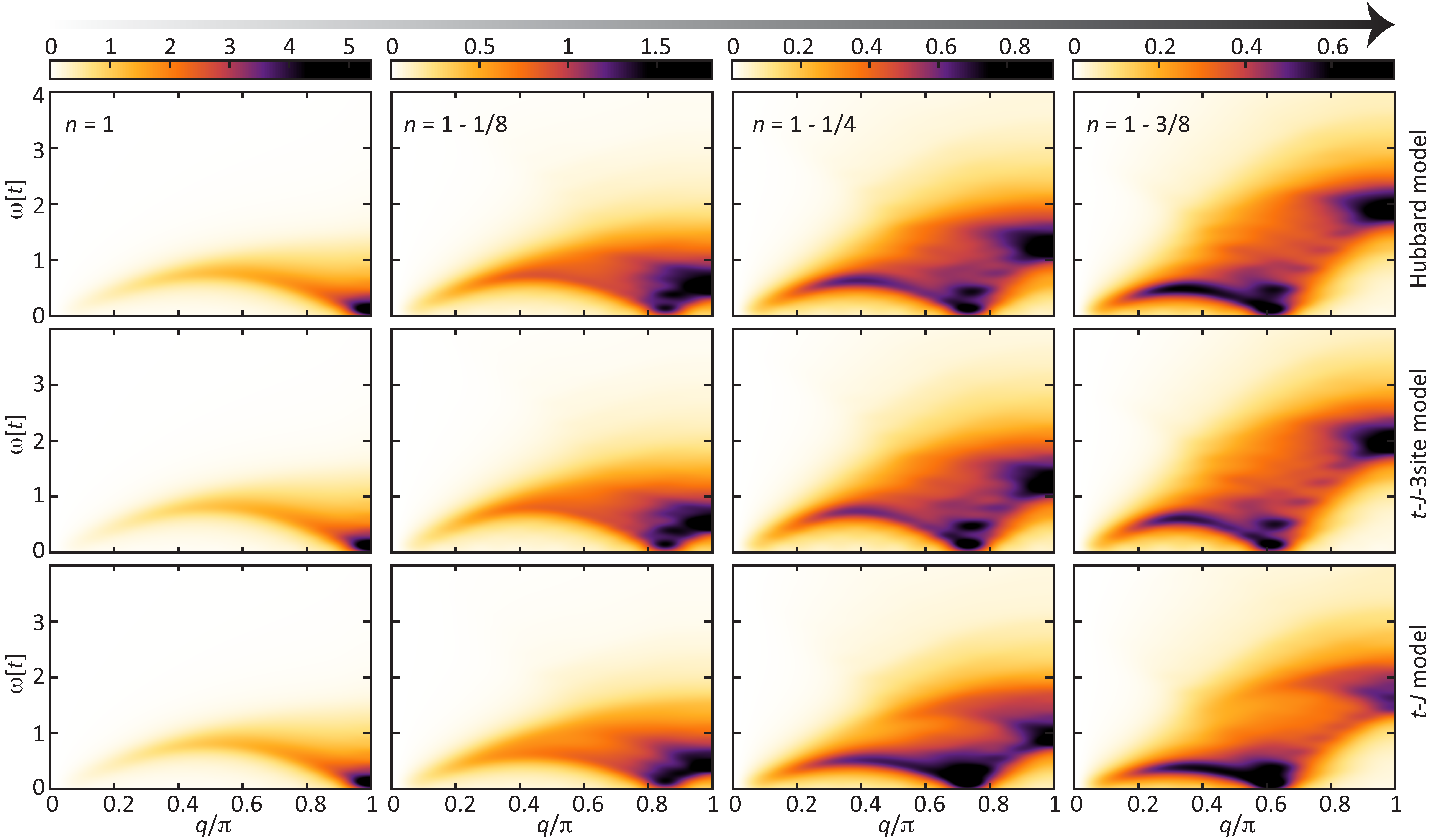}
\caption{$S(q,\omega)$ calculated using ED+CPT for various doping levels (left to right: number of electrons $n$=1, 0.875, 0.75, and 0.625) on a $16$-site 1D system with particle-hole symmetry ($t^\prime=0$). Upper panels -- Hubbard model with $U=8t$; middle panels -- \tJs\ model with $J=0.5t$; bottom panels -- \tJ\ model with $J=0.5t$. A Lorentzian broadening of $\delta = 0.15t$ is adopted in these calculations. Note that because of the particle-hole symmetry only the hole-doped spectrum is shown.
\label{fig:1}}
\end{center}
\end{figure*}

\subsection{Particle-Hole Symmetric System: $t^\prime=0$}
The system without $t^\prime$ exhibits particle-hole symmetry, so spin excitations are equivalent for hole and electron dopings. Figure~\ref{fig:1} shows $S(q,\omega)$ for the doping levels $x=0,\pm 1/8,\pm 1/4$ obtained by ED+CPT calculations. 
In the half-filled case ($n=1$, left panels of Fig.~\ref{fig:1}), $S(q,\omega)$  calculated for Hubbard, \tJs\, and \tJ\ models are almost identical. The spectrum shows gapless excitations at $q=0$ and $q=\pi$.
As there is only quasi-long-range AFM order, the spin spectrum does not exhibit sharp magnon excitations, but instead it displays a continuum between $\frac{\pi J}{2} \sin q$ and $\pi J \sin \frac{q}{2}$. This so-called two-spinon continuum stems from deconfined spin fractionalization: the $S=1$ spin excitation is constructed by two quasiparticles (spinons) carrying each $S=\frac12$ spin. A single $S=1$ spin flip then decays into an even number of spinons. With the quasi-long-range order, these spinons are correlated and concentrate mostly at the lower bound. In this case, due to the complete separation of spin and charge degrees of freedom (also will be shown later in the mean-field calculation), the spin models faithfully capture the spin texture of the Hubbard model.~\cite{Karbach1997,Bougourzi1998,kung2017}

Upon (either hole or electron) doping, the zero-mode momentum shifts from $\pi$ to smaller $q$ values. Due to a finite spectral broadening, it is hard to exactly extract the position of the gapless excitation, but it is close to $q=\pi(1-|x|)$, consistent with Refs.~\onlinecite{preuss1994spectral,NoceraPRB2016}. The lower limit of spin excitations slightly softens; in contrast, the upper limit increases with doping. At the same time, the spectral weight spreads out instead of concentrating on the lower bound, consistent with the destruction of the quasi-long-range order. Particularly, the system  maintains a large residual of the $q=\pi$ excitations, although being pushed to higher energies. 

Away from half filling, the spin models start to show a deviation from the Hubbard model (see middle to right panels of Fig.~\ref{fig:1}). Already seen at $x=0.125$ doping, the \tJ\ model underestimates the spin excitation energy, while the \tJs\ model slightly overestimates it. The 3-site term favors the persistence of spin correlations upon doping and helps to maintain a large curvature $\sim J$ in $S(q,\omega)$.~\cite{jia2014persistent, kung2017}
For the same reason, the \tJs\ model reproduces the spectral weight distribution of the Hubbard model better than \tJ\ at all doping levels. We note that even higher-order expansion terms may be important, which could account for the slight overestimation of spinon energy in the \tJs\ model.

\begin{figure*}[!t]
\begin{center}
\includegraphics[width=1.95\columnwidth]{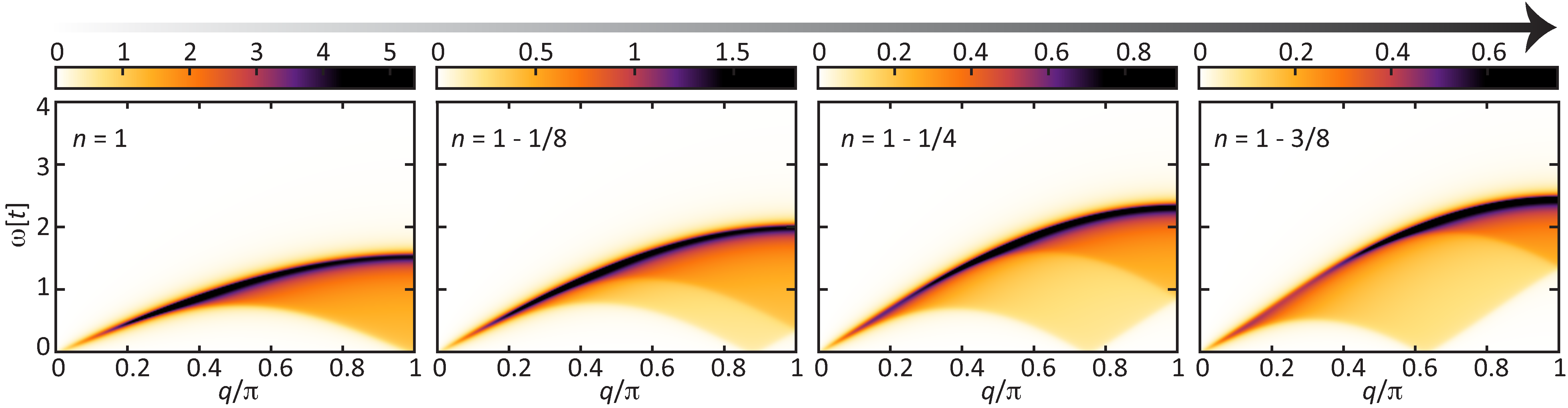}
\caption{$S(q,\omega)$ calculated using the slave-boson mean-field approach for the \tJ\ model with $J=0.5t$, $t^\prime=0$, at various doping levels (left to right: number of electrons $n$=1, 0.875, 0.75 and 0.625). 
A Lorentzian broadening of $\delta = 0.05t$ is adopted in these calculations. Note that because of the particle-hole symmetry only the hole-doped spectrum is shown.
\label{fig:2}}
\end{center}
\end{figure*}
\begin{figure*}[!ht]
\begin{center}
\includegraphics[width=2\columnwidth]{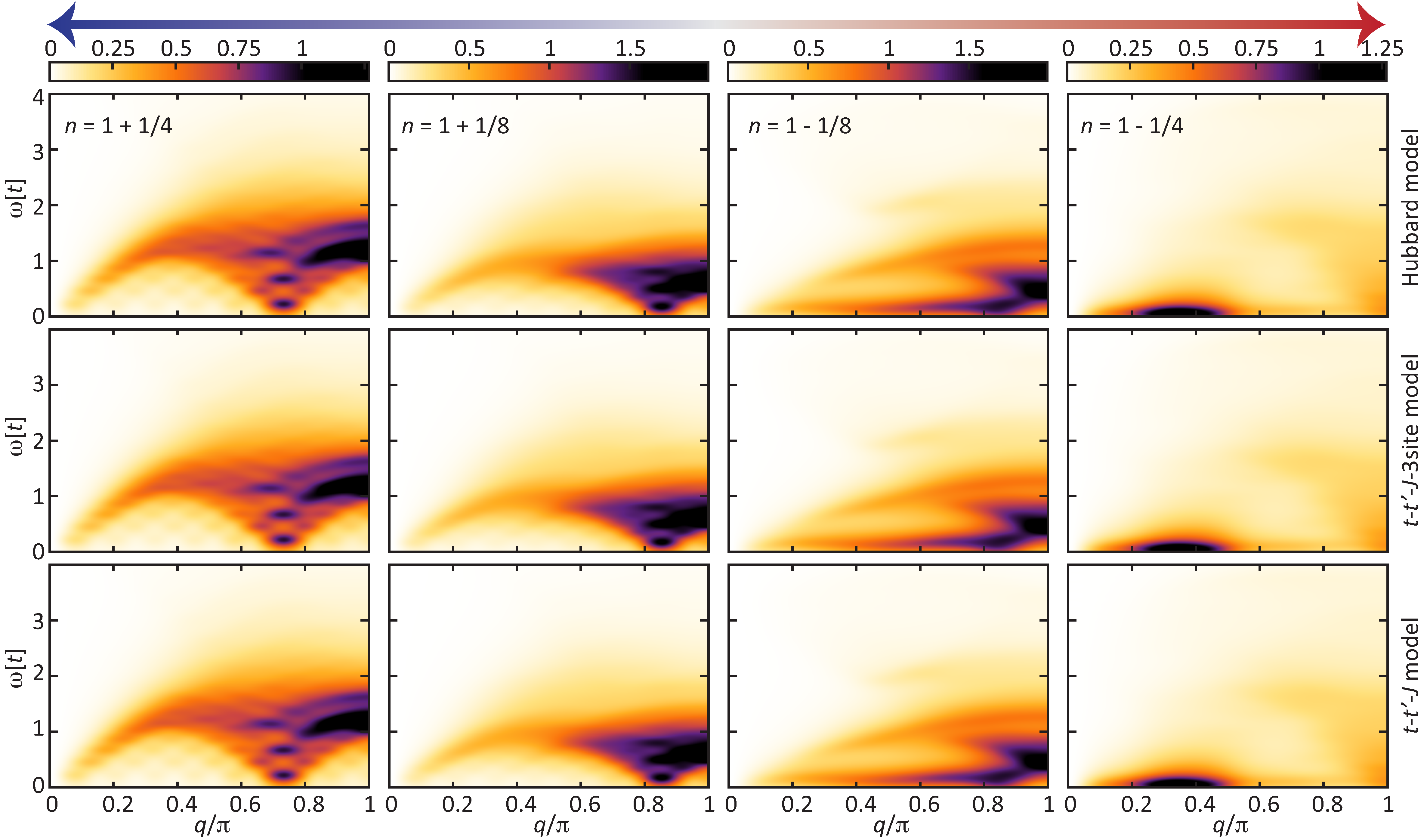}
\caption{
$S(q,\omega)$ calculated using ED+CPT for various doping levels (left to right: number of electrons $n$=1.25, 1.125, 0.875, 0.75) on a $16$-site 1D system without particle-hole symmetry ($t^\prime=-0.3t$). Upper panels -- Hubbard model with $U=8t$; middle panels -- \tJs\ model with $J=0.5t$; bottom panels -- \tJ\ model with $J=0.5t$. A Lorentzian broadening  of $\delta = 0.15t$ is adopted in these calculations.  
\label{fig:3}}
\end{center}
\end{figure*}

To help understand the doping dependence of spin excitations, we compute $S(q,\omega)$ of the \tJ\ model using the analytical slave-boson mean-field approach (see Fig.~\ref{fig:2}).
The compact supports of spin excitations (i.e. regions of allowed excitations) obtained in mean-field theory and in ED+CPT agree well, which means that the analytical approach can provide a theoretical framework to explain the observed doping evolution of spin excitations. We note, however, that the spectral weight obtained in mean-field theory does not carry any relevant information, so it should not be compared to the spectral weight obtained by ED+CPT. While more details  will be given in Sec.~\ref{sec:discussions}, we already note here that the the mean-field approach provides a straightforward explanation for the shift of the zero-mode momentum upon doping, which follows from particle-hole excitations between the two spinons on the Fermi surface separated by $2q_F = \pi(1-|x|)$.

\begin{figure*}[!ht]
\begin{center}
\includegraphics[width=2\columnwidth]{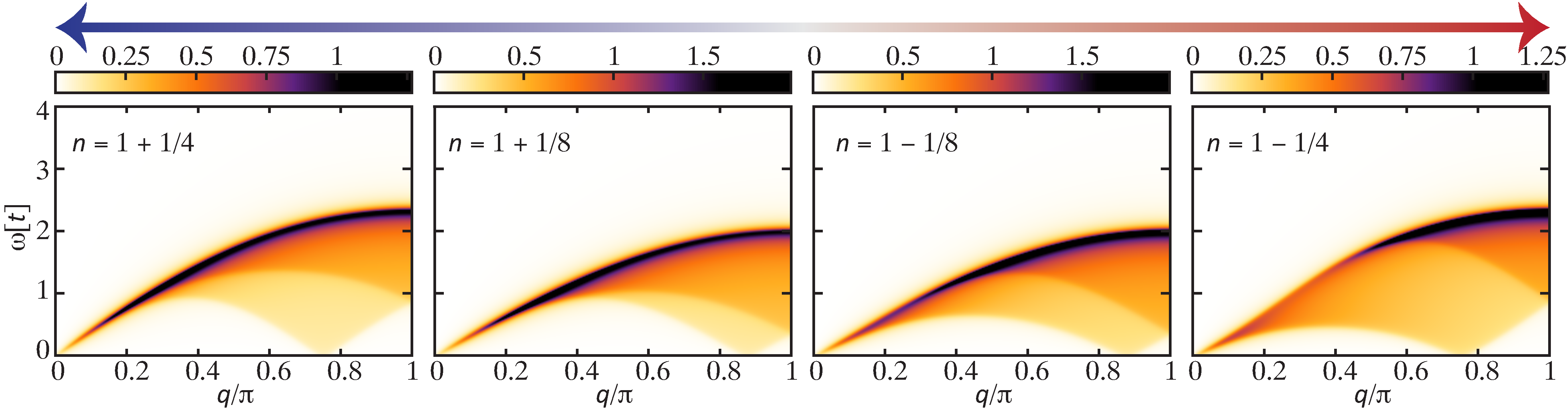}
\caption{$S(q,\omega)$ calculated using the slave-boson mean-field approach for the \tJ\ model with $J=0.5t$, $t^\prime=-0.3t$, at various doping levels (left to right: number of electrons $n$=1.25, 1.125, 0.875, 0.75). A Lorentzian broadening of $\delta = 0.05t$ is adopted in these calculations. 
\label{fig:4}}
\end{center}
\end{figure*}

\subsection{System Without Particle-Hole Symmetry: $t^\prime\neq 0$}
While the $t^\prime= 0$ case provides the simplest scenario to study the impact of carrier doping on spin excitations, it cannot be directly applied to a realistic material where the particle-hole symmetry is usually broken by further neighbor hopping terms. To understand systems with electron-hole asymmetry, we next examine the dynamical spin structure factors of the Hubbard model with $t^\prime\neq 0$, as well as for the \ttprJ  and \ttprJs\ models.

Figure~\ref{fig:3} shows $S(q,\omega)$ calculated by CPT+ED.\footnote{The unshown half-filled case is very similar to the $t^\prime = 0$ result in Fig.~\ref{fig:1}.} 
With a small $t^\prime = -0.3t$, the spectra already exhibit clear electron-hole asymmetry.
The electron-doped $S(q,\omega)$ displays an obvious hardening in energy. This is striking, because the curvature of the two-spinon continuum reflects the effective local spin-exchange interaction. When the system is doped away from half filling, the (quasi-)ordered AFM background is expected to be rapidly destroyed, which brings down the cost of a single spin flip. The fact that doping 
induces a hardening indicates that the effective interaction is strengthened, although the overall spectral weight drops compared to half filling due to the dilution of spin singlets. This persistence and hardening of spin excitations have been observed in 2D cuprate systems, both experimentally and numerically.~\cite{le2011intense,dean2013persistence,LeeNaturePhys2014,ishii2014high, jia2014persistent} The similarity between the 1D and 2D results suggests that the origin is generic to the model and does not depend on the system dimensionality. We attribute this phenomenon to a combined effect of the 3-site term and $t^\prime$ hopping (explained in detail in Sec.~\ref{sec:discussions}). The hardening maximizes at around $x=0.25$ electron doping; for even higher doping levels the carrier itineracy starts to dominate the dynamics and the spin correlations soften.


In contrast, with hole doping the dominant features of $S(q,\omega)$ exhibit a rapid softening (see right panels of Fig. 3). In 2D systems, however, the spin spectra show only little softening.~\cite{Dellea2017,dean2013persistence}
While the reason for such a strong quantitative difference is not exactly clear, we suggest that in general it stems from the well-known distinct nature of collective excitations in undoped 1D and 2D systems.
Interestingly, such a strong softening of the two-spinon continuum observed here for the $t^\prime=-0.3t$ case also contrasts with the 
four-legged \ttprJ\ ladders~\cite{white1999}, which behave similarly to the 2D system with $t^\prime=-0.25t$.~\cite{Tohyama1999}

At the same time, the overall spectral weight of $S(q,\omega)$ for the hole-doped chain is smaller compared to that at the corresponding electron doping. This is because $t^\prime$ plays an opposite role here: it weakens the effective spin-exchange energy of collective excitations and helps to destroy the AFM background. 
Because of that, the spectra of the spin models differ from that of the Hubbard model. Interestingly, in all models, the higher energy features near $q=\pi$ are similar for both electron and hole dopings, 
which could probably be attributed to the equal spin-exchange interaction $\sim J$ describing local (not collective) excitations.

With finite $t^\prime$, the slave-boson mean-field theory also can qualitatively capture the compact support upon hole- and electron-doping the \tJ\ model (see Fig.~\ref{fig:4}): Indeed, the lower bound of two-spinon continuum hardens for electron doping and softens for hole doping. We note that the curvatures of the two separate branches at the lower bounds of the two-spinon continuum for finite doping become different when $t^\prime\neq 0$. This reflects the distinct motion of spinon and anti-spinon due to the absence of particle-hole symmetry. The upper bound and the nesting momentum of the two-spinon continuum are sensitive only to the absolute value of the doping concentration $|x|$. This reveals the fixed Fermi geometry and bandwidth, as will be discussed in Sec.~\ref{sec:discussions}.  

\section{Discussion}\label{sec:discussions}
\begin{figure}[!t]
\begin{center}
\includegraphics[width=0.8\columnwidth]{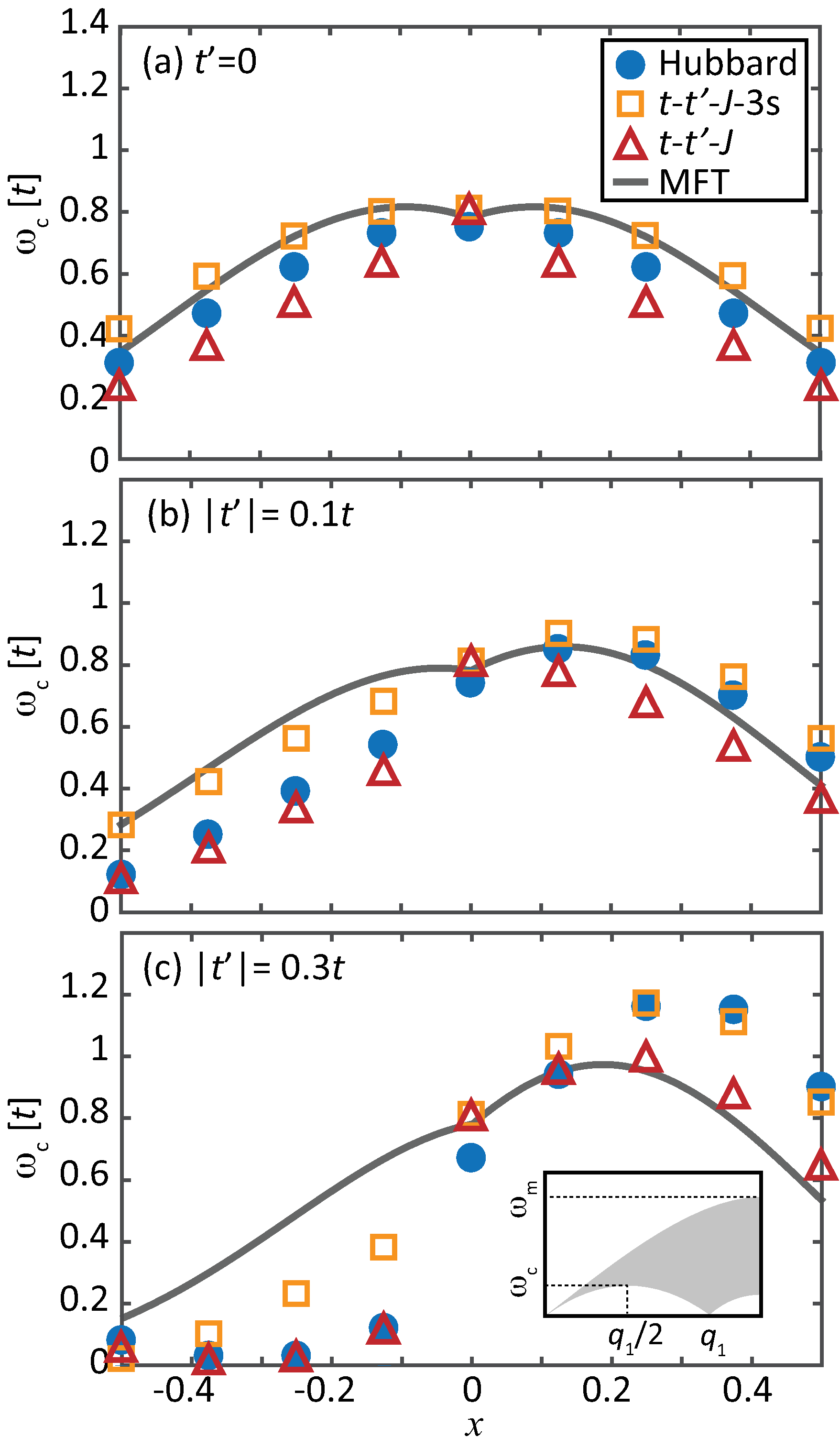}
\caption{The characteristic spin excitation energy $\omega_c$ (see text) as a function of doping $x$, obtained from ED+CPT results for the Hubbard model (solid circles), \tJs\ model (open squares), and \tJ\ model (open triangles) with different values of $t^\prime$.
Negative (positive) $x$ corresponds to hole (electron) doping.
The gray curve denotes the corresponding mean-field fermionic $\xi(k)$ bandwidth. The inset of panel (c) illustrates the definition of $\omega_c$ and $\omega_m$ as explained in the text.
\label{fig:6}}
\end{center}
\end{figure}

In this section, we proceed with more quantitative analysis of spin excitations presented in Sec.~\ref{sec:results}. To parameterize the characteristic energy of spin excitations obtained by ED+CPT, we define the peak energy at $q=q_1/2$ as $\omega_c$ (i.e. the maximum energy in the lower bound of the two-spinon continuum), where $q_1$ corresponds to the zero-mode momentum. We also define the maximum spin excitation energy at $q=\pi$ as $\omega_m$ [see Fig.\ref{fig:6}(c) inset].
In the mean-field picture, $\omega_c$ and $\omega_m$ correspond to the  bandwidth of filled spinons and the full band dispersion.
Using these parameters, we quantitatively examine the consistencies and discrepancies among different models and methods. We then utilize the straightforward mean-field picture to provide intuitive explanations of the features observed in numerical calculations.

\subsection{Quantitative Comparison between Different Models and Methods}
Using the results presented in the previous section, we extract the evolution of $\omega_c$ upon electron and hole doping for the Hubbard, \tJ\, and \tJs\ model with different values of $t^\prime$, as shown in Fig.~\ref{fig:6}.
Concentrating first at the $t^\prime=0$ case, we note that the $\omega_c$ values calculated using different models match well at half filling and follow the same trend with doping. Interestingly, the \tJ\ model underestimates the spin excitation energy, while the \tJs\ overestimates it by a similar extent. The underestimation in the \tJ\ model takes place as one neglects the 3-site term, which lowers the kinetic energy by protecting the AFM sublattice and favors spin order. The overestimation in the \tJs\ model might arise as the model is projected onto a restricted spin Hilbert space, which
suppresses charge fluctuations.
The quantitative mismatch demonstrates that \tJl\ models
cannot exactly reproduce spin excitations of the Hubbard model even up to the second order $t/U$ expansion. This difference is not obvious at half filling due to the frozen charge degrees of freedom, but becomes more pronounced with doping due to enhanced charge fluctuations.

A similar trend also exists for finite $t^\prime$, as shown in Figs.~\ref{fig:6}(b) and (c). The $\omega_c$ obtained from {\ttprJ} and {\ttprJs} models are respectively lower and higher than that from the Hubbard model. Spin excitations calculated for all three models exhibit a hardening on the electron-doped side, and a softening on the hole-doped side. Microscopically, this is a direct consequence of the further-neighbor hopping $t^\prime$, which breaks the particle-hole symmetry by enhancing (suppressing) the tunneling of holes (electrons). Thus, for electron doping, it helps to decrease the scrambling of spin order induced by the motion of doped electrons. For hole doping, on the contrary, $t^\prime$ promotes motion of the doped holes into ordered spin clusters hence destroying the AFM order. Therefore, we see a rapid drop of spin excitation energy (and the spectral weight) with hole doping but no obvious change with electron doping.

Interestingly, the Hubbard model results are closer to the \ttprJ\ model on the hole-doped side and closer to the \ttprJs\ model on the electron-doped side. The former can be easily understood as the 
\ttprJ\ model includes the complete second-order $t/U$ expansion, while the latter can be attributed to the fact that the \ttprJs\ model 
misses the longer-range 3-site terms $\propto t^\prime t/U$. We note that the 3-site term, in general, helps maintaining the spin order and enhancing the spin excitation energy (reflected by the difference of \ttprJ\ and \ttprJs\ models in Fig.~\ref{fig:6}), while the longer-range 3-site terms can suppress spin correlations upon hole doping. It is the combined effect of $t^\prime$ and 3-site term that leads to the hardening of $S(q,\omega)$ on the electron-doped side. As a consequence, we observe that the critical doping -- the $x$ where spin excitation hardens the most -- depends monotonically on $t^\prime$. This relation provides a way to experimentally estimate the physical parameter $t^\prime$.

Finally, we compare the numerical $\omega_c$ from ED+CPT with that from the slave-boson mean-field theory, which is shown as gray curves in Fig.~\ref{fig:6}. The mean-field result faithfully reflects the correct $\omega_c$ when $t^\prime=0$. However, although the overall trend is correct, it is quantitatively different from the numerical value of any model. Specifically, the mean-field calculation underestimates the impact of $t^\prime$, which can be attributed to two reasons. First, the derivation of Eq.~\eqref{eq:xik} considers the $t^\prime$ impact only as a hopping term and ignores its indirect impact on the spinon configurations (i.e. correction to $J$). 
Second, Fig.~\ref{fig:6} compares the lower bound of the mean-field two-spinon continuum with the peak position of $S(q,\omega)$ computed by ED+CPT. While the spectral weight is expected to concentrate mostly on its lower bound at half filling, a redistribution happens with doping. It is clear that in electron-doped systems (see the left column of Fig.~\ref{fig:3}), the spectral weight spreads out towards higher energy. Therefore, the numerically extracted peak position $\omega_c$ is not necessarily located at the lower bound of the compact support. The fact that the mean-field result deviates more from the numerical one in hole-doped systems 
indicates that the mean-field ansatz is violated at large hole doping.

\begin{figure}[!ht]
\begin{center}
\includegraphics[width=\columnwidth]{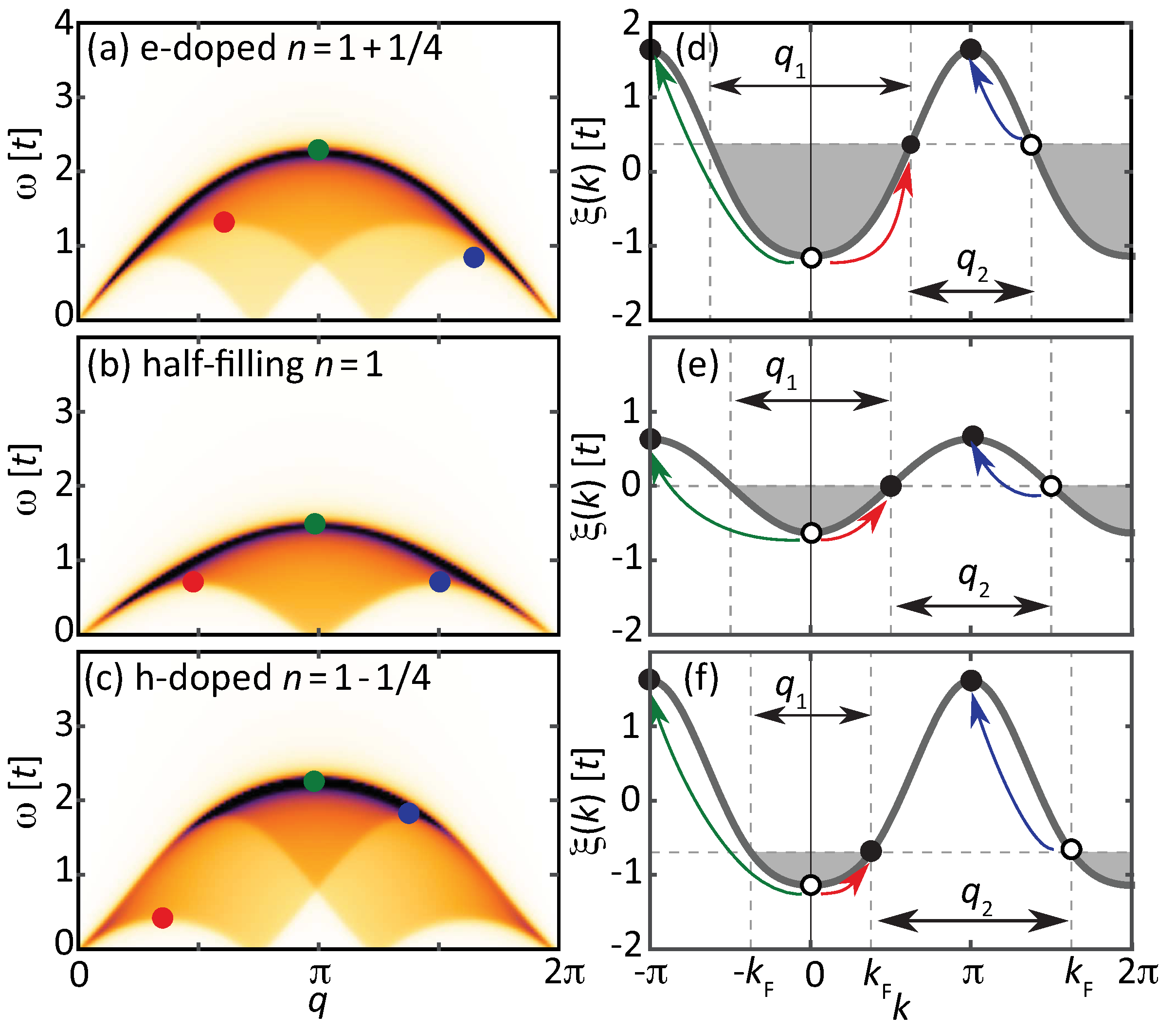}
\caption{(a-c) $S(q,\omega)$ calculated using the slave-boson mean-field approach for the \tJ\ model with $J=0.5t$, $t^\prime=-0.3t$, at the doping level $n=1.25$, 1.0 and 0.75, respectively. (d-f) Mean-field fermionic spinon band $\xi_k$ Eq.~\eqref{eq:xik} for the same doping levels. The shaded area of the fermionic band $\xi_k$ is occupied by particles (spinons), while the gray dashed lines denote the Fermi energy $E_F$ and momenta $k_F$. 
The vectors $q_1$ and $q_2$ denote the allowed momenta for zero-energy particle-hole excitations (see text).
The colored (red, green, or blue) arrows in (d-f) indicate the particle-hole excitations that contribute to the excitations at particular ($q, \omega$) points denoted by the corresponding colored dots in (a)-(c).
\label{fig:5}}
\end{center}
\end{figure}

\subsection{Insights from the Slave-Boson Mean-Field Theory}
Within the slave-boson mean-field theory, the spin dynamical structure factor $S(q,\omega)$ is completely governed by the spinon dispersion $\xi_k$ in Eq.~\eqref{eq:xik} through the Lindhard susceptibility. Therefore, the doping evolution of $S(q,\omega)$ can be understood intuitively by the corresponding noninteracting spinon bandstructure. Figure~\ref{fig:5}(left panels) shows the $S(q,\omega)$,
calculated in this mean-field picture,
 of electron-doped, undoped, and hole-doped systems in a full Brillouin zone. 

First, we observe that zero-energy excitations in $S(q,\omega)$ originate from particle-hole (\textit{i.e.}~spinon-antispinon) excitations on the spinon Fermi surface.
Irrespective of the doping level, in the full Brillouin zone such a zero-energy scattering is allowed only at two different nesting momenta $q_1=2k_F$ and $q_2=2\pi-2k_F$.
Since the spinon band dispersion is convex, the Fermi momentum $k_F$ is independent of the hopping parameters, \textit{i.e.}~$k_F={\pi(1-|x|)}/{2}$ (see Sec.~\ref{method:MFT}). Therefore, $q_1$ and $q_2$ depend only on the absolute value of doping concentration $|x|$. 
This accounts for the linear shift of zero-energy momentum with doping (see Figs.~\ref{fig:2} and \ref{fig:4}), and explains their independence of the sign of $x$.

Second, the two well-visible branches of the two-spinon continuum can be tracked down to the scattering of spinons to the Fermi surface (red arrows), and from the Fermi surface (blue arrows). These two scattering channels are governed by the band structure below  $E_F$ (for spinons) and above $E_F$ (for antispinons), respectively. Without $t^\prime$, the particle-hole symmetric system has a symmetric $\xi_k$ below and above $E_F$, resulting in the same curvatures of the two branches in Fig.~\ref{fig:2}. With a finite $t^\prime$, the scattering of spinons and of antispinons is allowed in two distinct parts of $\xi_k$, leading to different effective masses of the two-spinon excitations. Thus, spinons become heavier at hole-doping, while antispinons become lighter (also see Fig.~\ref{fig:4}).

Third, the noninteracting band picture helps to understand the origin of the hardening (softening) of $S(q,\omega)$ upon electron (hole) doping with finite $t^\prime$. With a negative $t^\prime$, the band structure $\xi_k$ suffers from a doping-induced `unbalance' (due to the $\cos 2k$ term in Eq.~\eqref{eq:xik}) below and above $\omega=0$ (Fermi level at half filling). The band below $\omega=0$ is flatter with a narrow bandwidth, while sharper above it. Therefore, the increase of the depth of spinon Fermi sea compensates the band flattening, leading to an unchanged or even hardened two-spinon excitations. On the other hand, hole doping dries out the Fermi sea and flattens the band bottom, both of which lead to a softening of two-spinon excitations.

Finally, the upper edge $\omega_m$ of the two-spinon continuum is determined by the excitation from the bottom to the top of the spinon band. 
Since in Eq.~\eqref{eq:xik} $t^\prime$ appears in front of $\cos 2k$, it does not change the energy difference between the top ($k=\pi$) and the bottom ($k=0$) of the spinon band. Therefore, $\omega_m$ remains the same with and without $t^\prime$, for both hole and electron doping (compare Figs.~\ref{fig:2} and ~\ref{fig:4}). Moreover, as the total bandwidth of $\xi(k)$ increases with doping, $\omega_m$ always increases with doping $|x|$ away from half filling.

\section{Conclusions}\label{sec:conclusions}
We have systematically studied the doping evolution of spin excitations in 1D Hubbard, \tJ\, and \tJs\ models, via both numerical and analytical approaches.
Their spectra exhibit several common trends, including (i) the shift of the zero-energy momentum with doping, (ii) a relatively weak softening of the dominant feature upon doping a particle-hole symmetric system, (iii) a weak hardening (strong softening) of the dominant feature upon electron (hole) doping a system with finite longer-range hopping
$t^\prime$. At a quantitative level, however, spin excitations from both \tJ\, and \tJs\ models differ from that of the Hubbard model. In particular, for the hole-doped system with finite $t^\prime$, the \tJ\ model seems to be closer to the Hubbard model than the \tJs\ model. This counterintuitive result suggests that to obtain a genuine quantitative agreement between the Hubbard and a low-energy spin model, one needs to supplement the latter by higher-order or longer-range terms in the expansion in $t/U$ and $t^\prime/U$ of the original Hubbard model.

In order to qualitatively understand the changes in the spin spectrum upon doping, we compare the numerical results with that from a slave-boson mean-field theory for the \tJ\ model.At half filling the latter faithfully captures the compact support of $S(q,\omega)$. Moreover, it provides a straightforward picture to understand the doping evolution of nesting momentum, as well as the hardening and softening of the two-spinon continuum upon doping.
The hardening and softening of spin excitations observed in the doped 1D models are comparable to those observed experimentally in 2D cuprates. Although the tendency towards the persistence or hardening is far weaker in 1D, it provides a simpler platform to better understand these phenomena. We note that the role of the 3-site terms are far more important in 2D than in 1D, since in 2D the 3-site terms allow for both next-nearest and third-neighbor hoppings.

The quantitative analysis using different models and approaches enriches our understanding of spin excitations of the 1D Hubbard model. The nature of hardening or softening in $S(q,\omega)$ reflects the underlying competition between spin order and carrier itinerancy, and more importantly, the interplay among spin-exchange interaction, longer-range hopping, and correlated 3-site term.
The presented results should be broadly applicable to inelastic neutron and x-ray scattering experiments on quasi-1D correlated systems -- such as doped cuprates and 1D cold atom systems.

\begin{acknowledgments}
We are grateful to Clement Bazin for his initial work on this problem.
Y.~W. acknowledges the Postdoctoral Fellowship in Quantum Science of the Harvard-MPQ Center for Quantum Optics and AFOSR-MURI Quantum Phases of Matter (Grant FA9550-14-1-0035). B.~M. and T.~P.~D. acknowledges support from the U.S. Department of Energy, Office of Science, Office of Basic Energy Sciences, Division of Materials Sciences and Engineering, under Contract No. DE-AC02-76SF00515. K.~W. acknowledges support from Narodowe Centrum Nauki (NCN), under Project No.~2016/22/E/ST3/00560 and Project No.~2016/23/B/ST3/00839. This research used resources of the National Energy Research Scientific Computing Center (NERSC), a U.~S. Department of Energy Office of Science User Facility operated under Contract No. DE-AC02-05CH11231.
\end{acknowledgments}


\end{document}